\documentclass[journal]{IEEEtran}
\usepackage{amsmath,amsfonts}
\usepackage{algorithmic}
\usepackage{array}
\usepackage[caption=false,font=normalsize,labelfont=sf,textfont=sf]{subfig}
\usepackage{textcomp}
\usepackage{stfloats}
\usepackage{url}
\usepackage{verbatim}
\usepackage{graphicx}
\usepackage{orcidlink}
\def\BibTeX{{\rm B\kern-.05em{\sc i\kern-.025em b}\kern-.08em
    T\kern-.1667em\lower.7ex\hbox{E}\kern-.125emX}}
\usepackage{balance}

\usepackage{xspace}     
\usepackage{xpunctuate} 
\usepackage{dsfont}
\usepackage{enumitem}   

\usepackage{adjustbox}
\usepackage{subcaption} 
\usepackage{array}      
\usepackage{tikz}       
\usepackage{hyperref}
\usepackage{booktabs} 

\newcommand{\ie}{{i.e.,}\xspace}
\newcommand{\eg}{{e.g.,}\xspace}
\newcommand{\etal}{{et~al\xperiod}\xspace}

\newcommand{\etc}{{etc\xperiod}\xspace}  

\newcommand{\insightText}[1]{\emph{``#1''}} 

\newcommand{\myquote}[1]{\emph{``#1''}} 
\newcommand{\myquoteparticipant}[2]{    
    \myquote{#1}~(#2)}

\newcommand{\parHeading}[1]{\vspace{4px}\noindent{\textbf{#1}}}
\newcommand{\parHeadingNoSpace}[1]{\noindent{\textbf{#1}}}

\newcommand{\1}[1]{\mathds{1}\left[#1\right]} 

\usepackage{xargs}      
\usepackage{soul}       
\usepackage{color}      

\usepackage{caption}

\usepackage{subcaption}
\usepackage{relsize} 

\definecolor{LIGHTPINK}{RGB}{237,157,202}
\definecolor{LIGHTRED}{RGB}{210,121,121}
\definecolor{LIGHTORANGE}{RGB}{230,170,50}
\definecolor{LIGHTGOLD}{RGB}{210,194,121}
\definecolor{LIGHTGREEN}{RGB}{121,210,121}
\definecolor{LIGHTAQUA}{RGB}{121,206,210}
\definecolor{LIGHTBLUE}{RGB}{121,124,210}
\definecolor{LIGHTPURPLE}{RGB}{153,102,255}
\definecolor{RED}{RGB}{178,34,34}
\definecolor{GRAY}{RGB}{166,166,166}
\definecolor{WHITE}{RGB}{255,255,255}

\newif\ifshowcomments
\showcommentstrue

\newcommand{\todo}[1]{
    \ifshowcomments
    \addcontentsline{toc}{subsection}{  %
        \protect\numberline{}           
        \textcolor{RED}{[TODO] #1}}     
    \textcolor{RED}{([TODO] \emph{#1})}
    \fi}  

\newcommandx{\jane}[2][1=] 
    {\ifshowcomments \setulcolor{LIGHTGREEN}{\ul{#1}} \textcolor{LIGHTGREEN}
    {[\textbf{Jane:} #2]}\fi}

\newcommandx{\junran}[2][1=] 
    {\ifshowcomments \setulcolor{blue}{\ul{#1}} \textcolor{blue}
    {[\textbf{Junran:} #2]}\fi}

\newcommandx{\leilani}[2][1=] 
    {\ifshowcomments \setulcolor{blue}{\ul{#1}} \textcolor{LIGHTPURPLE}
    {[\textbf{Leilani:} #2]}\fi}
    
\newcommandx{\guest}[3][1=]
    {\ifshowcomments \setulcolor{LIGHTORANGE}{\ul{#1}} \textcolor{LIGHTORANGE} 
    {}\fi}

\newcommand{\existingchart}[1]{$e_{#1}$}
\newcommand{\addedchart}[1]{$a_{#1}$} 

\newcommand{\likert}{
\begin{figure}[t]
    \centering
    \includegraphics[width=\columnwidth]{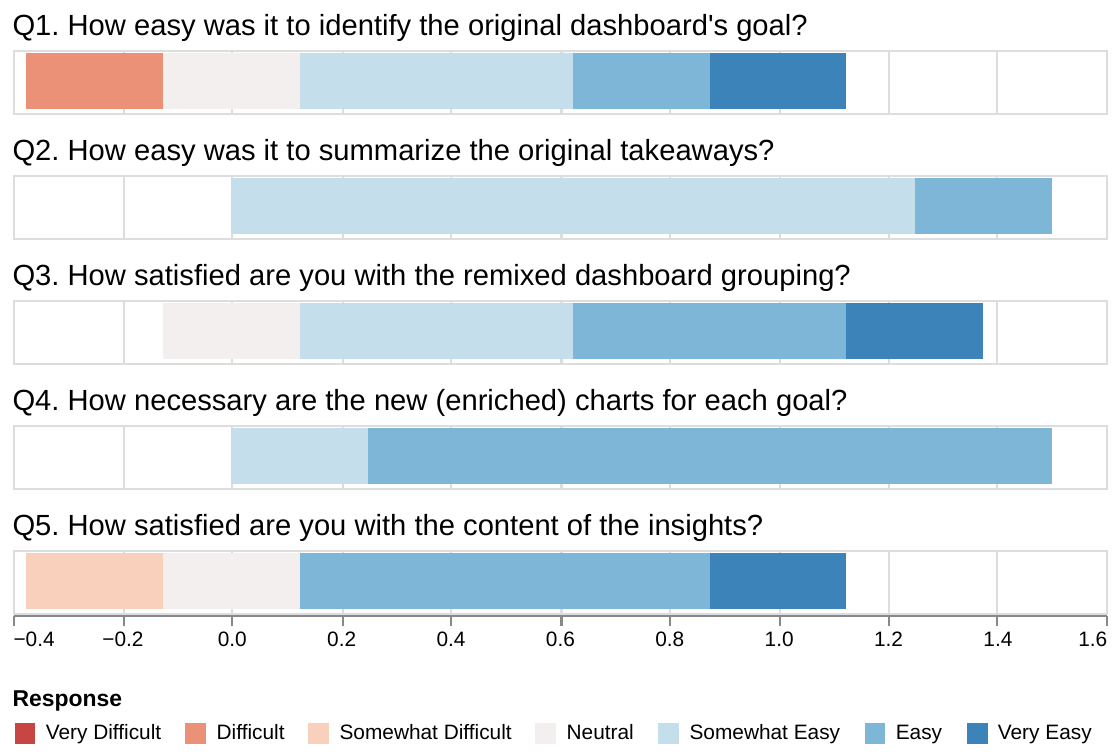}
    \caption{Six experts rated the input dashboard and components of our pipeline including grouping, enrichment, and insights. Note: the scale used difficult / unsatisfied / unnecessary versus easy / satisfied / necessary for the ratings, as appropriate.}
    \label{fig:likert}
\end{figure}
}

\newcommand{\pipeline}{
\begin{figure*}[t]
    \centering
    \includegraphics[width=\textwidth]{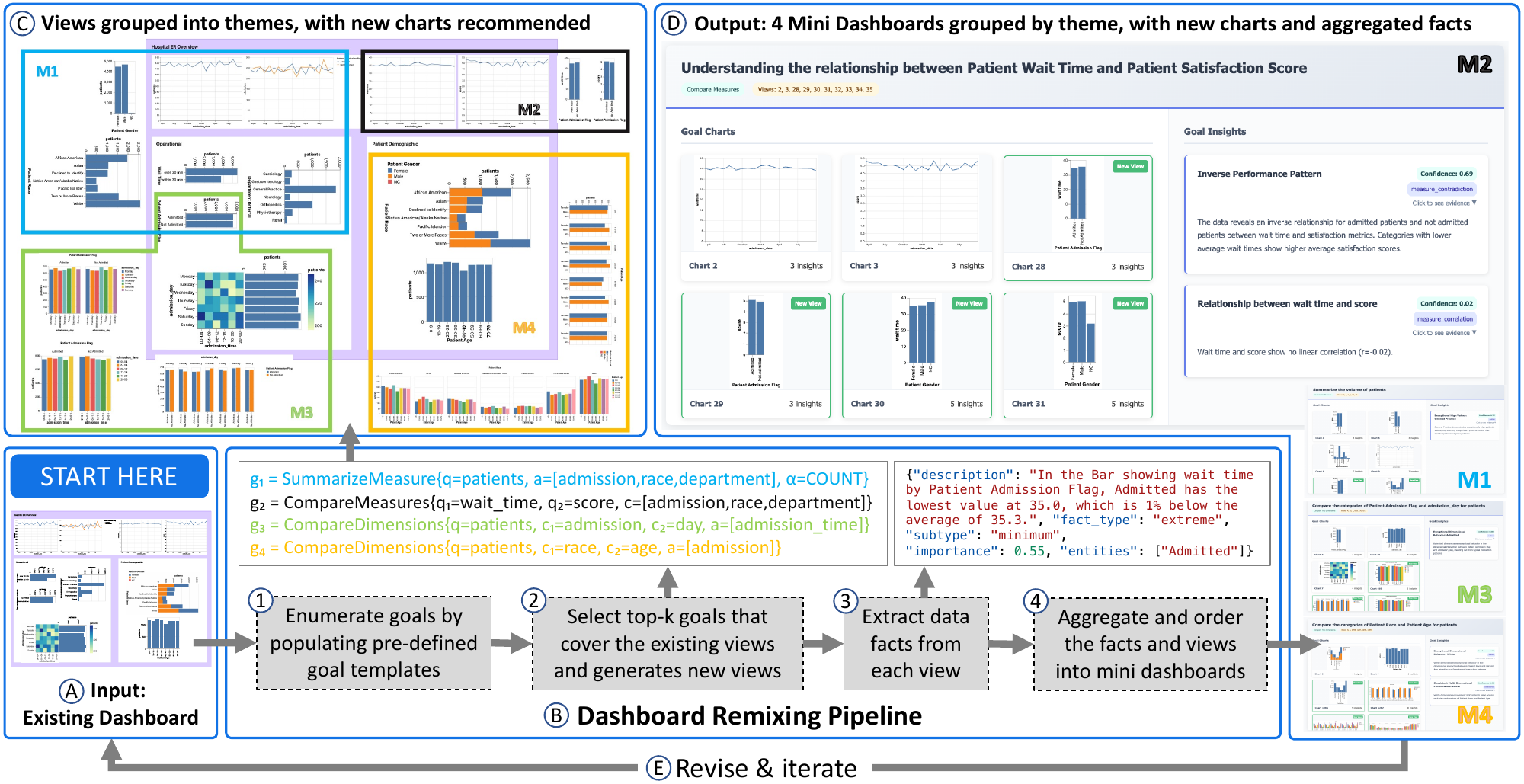}
    \caption{Our recommendation pipeline \textcircled{\footnotesize B} consists of four steps (\textcircled{\footnotesize 1} - \textcircled{\footnotesize 4}). The input \textcircled{\footnotesize A} is a dashboard with its chart specification and underlying data tables. The output \textcircled{\footnotesize D} is a structured webpage consisting of view groups, descriptions of the goals, and data insights reflecting the goals and views in each group. \textcircled{\footnotesize C} is an illustration of existing and added views grouped by goals. 
    }
    \label{fig:pipeline}
\end{figure*}

}

\begin{document}

\title{Dashboard Remixing: Goal-oriented Organization and Enrichment of Visualizations and Insights}
\author{Junran Yang\,\orcidlink{0000-0002-8467-2917}, Leilani Battle\,\orcidlink{0000-0003-3870-636X}, and Jane Hoffswell\,\orcidlink{0000-0002-9871-4575}
\thanks{Junran Yang and Leilani Battle are with Paul G. Allen School of Computer Science and Engineering at the University of Washington (e-mail: junran@cs.washington.edu; leibatt@cs.washington.edu). Jane Hoffswell is with Adobe Research (e-mail: jhoffs@adobe.com).}}


\maketitle

\begin{abstract}
Dashboard creation requires iteratively authoring multiple views and organizing them into a structured format. 
However, creators often struggle to step back and assess their own work for organization and refinement due to cognitive limitations. 
In this work, we propose a pipeline to provide computational support for self-reflection during the design process. Our pipeline automatically infers analytical objectives, discovers implicit patterns, and structures those patterns to surface coherent analytical pathways.
It further enriches existing dashboard designs by generating complementary visualizations and aggregating natural language insights based on relationships across data entities (\eg~contrasting, drill-down, and relatedness) to reveal chart connections. 
Our rule-based pipeline provides objective, verifiable feedback, offering an external perspective while enabling creators to tell their own data stories. 
Four case studies and six expert interviews demonstrate how this approach helps creators reflect on organizational alternatives and surface hidden analytical opportunities in a principled way, reducing the cognitive burden while encouraging deeper analytical exploration.


\end{abstract}



\section{Introduction}
\label{sec:introduction}


Dashboards are essential for exploring and monitoring data as well as communicating insights to stakeholders for further decision-making. 
Authoring experiences typically involve first accumulating individual charts before organizing them into a structured dashboard~\cite{epperson2023declarative, battle2019characterizing}. 
Dashboard creators also iterate \textit{between} accumulating and organizing charts~\cite{froese2016lessons} as they refine their analytical focus to effectively represent the most important information. 
Current chart recommendation systems and dashboard authoring tools (\eg{} Tableau or Power~BI)
support the accumulation phase well by facilitating rapid chart prototyping.
However, creators often must step back from the data analysis process to assess the broader structure of their work. 
This reflection requires creators to {simultaneously identify the overarching themes, minimize redundancy to keep the analysis focused, and uncover gaps in the chart collection.}
However, each of these tasks poses significant cognitive challenges. Limited working memory~\cite{pirolli2005sensemaking} constrains how much structure and evidence creators can hold in mind simultaneously; recognizing cross-view patterns requires tracking relationships across scattered evidence at once; identifying gaps requires exhaustive comparison of attribute combinations against existing views; and cognitive fixation~\cite{bako2025unveiling} discourages exploration of organizational alternatives beyond initial intuitions. These demands highlight the difficulty in managing scale beyond creating individual charts and the urgency of stepping back to gain perspective on an evolving collection.

While specific tools for dashboard recommendation have advanced significantly~\cite{deng_dashbot_2022, lin2024dminer}, they mainly focus on generating individual charts to compose the dashboards (accumulation).
On the other hand, some prior work~\cite{pandey_medley_2022, shi_calliope_2021, wang_datashot_2020} addresses aspects of the organization after extracting fact-based insights. 
However, these approaches either assume the organizational structure is already determined~\cite{epperson2023declarative, pandey_medley_2022} or focus on progressing a storyline~\cite{shi_calliope_2021, wang_datashot_2020, zhao2023ChartStory}. 
Thus, there remains a gap: no existing tools provide computational support to help creators reflect on the dashboard design, such as identifying what themes exist, what charts are missing as evidential facts to support the theme, and what patterns connect multiple charts. 

In this work, we thus contribute a pipeline that provides such computational support for self-reflection and organization in iterations, with a dashboard or collection of analytical charts as input. 
Our approach involves four steps:
first, we extract analytical goals from existing views using a predefined taxonomy of dashboard objectives (e.g., \emph{``understand temporal patterns''} or 
\emph{``compare categorical values''}). 
We then identify a minimal set of themes that cover the dashboard's analytical scope using weighted set cover optimization and mixed integer linear programming (MILP). 
This step typically identifies two to four themes that best characterize what the dashboard investigates by considering all views holistically.
We then enrich each theme by generating new views that complement the existing analysis and objective. 
Next, we extract data facts using existing heuristics~\cite{ding2019quickinsights} and synthesize cross-view patterns using a semantic network that connects related insights across views. To synthesize the insights, we identify three relationship types (contrast, drill-down, and relatedness), and aggregate connected facts into text summaries describing the patterns within each theme.
The output is a structured dashboard with explicit themes, grouped and enriched views, and text summaries to provide creators with an external 
perspective on their exploratory work for review and refinement. 
Creators can thus 
compare their intended design goals to the pipeline outputs and iterate accordingly to bring the dashboard design and intended analytical focus in closer alignment.
Our pipeline uses rule-based, deterministic algorithms by default (with opt-in LLM enhancement), providing outputs that are verifiable and traceable to specific views and data. 
We further discuss integration with LLMs for richer semantic and domain interpretation or more flexible generation in Section ~\ref{sec:system}. 

{We distinguish between a dashboard's \textit{intended} goals held privately by its creator and its \textit{expressed} goals the arranged views communicate to a viewer. Our pipeline surfaces the latter, and comparing the two is what enables creator reflection. }
Our pipeline serves as a design reflection tool that provides creators with an external perspective on their evolving work for review and refinement. 
Rather than producing a final dashboard, the pipeline offers a computational viewpoint that creators can evaluate against their own intentions. 
Creators interact with the pipeline iteratively: they run it on a dashboard, introspect whether the identified themes align with their goals, bookmark the charts they want to retain to refine the design, and re-run as needed. 
The pipeline can point out that \emph{``these views all address temporal patterns''} or \emph{``you're missing a comparison across these dimensions,''} to offer a new perspective on analytical structure and coverage that complements creators' own judgment, which eases rather than replaces expert evaluation. 
Importantly, the structured output maintains the exploratory nature of dashboards such that it merges evidence to surface patterns without prescribing specific conclusions, enabling inspection of organized facts.

We present four real-world case studies based on dashboards from Tableau Public and the Power BI gallery (Section~\ref{sec:usecase}) 
{and conduct a study with six dashboard experts from across industry and academia (Section~\ref{sec:interview}), in which each expert first formed their own reading of a dashboard's goals and then compared the pipeline's output against it.}
{Experts praised the time efficiency and navigational benefits of goal identification and view reorganization as well as the pipeline's ability to surface hidden data patterns.
Experts also found the insight summarization valuable, but emphasized a continued need for new insight mining strategies.
Overall, the evaluation shows promise for enhancing dashboard interpretation and design.}

In summary, this paper makes the following contributions:
\begin{enumerate}
    \item a pipeline to enrich and restructure dashboards based~on inferred goals, helping creators refine their design; 

    \item a goal inference and optimization method that identifies minimal coherent dashboard themes, and generates new views to fill gaps in the dashboard's identified themes; 
    
    \item an insight aggregation method leveraging a semantic network to capture cross-view patterns; and
    
    \item case studies and expert interviews to collect feedback and demonstrate the effectiveness as a design reflection aid.

\end{enumerate} 


\section{Related Work}
\label{sec:relatedwork}
This work is motivated by research on dashboard recommendation, analytic communication, and automated insights.
\vspace{-.8cm}
\subsection{Dashboard Composition and Layout}
Foundational work has characterized what makes dashboards effective: Sarikaya~\etal~\cite{Sarkaya2019WhatDW} analyzed dashboard characteristics and usage contexts; Bach~\etal~\cite{Bach2022DashboardDP} examined design factors; Srinivasan~\etal~\cite{srinivasan2024dashboard} modeled composition and interaction patterns across a large Tableau Public corpus; and Tory~\etal{}~\cite{tory_finding_2023} studied challenges for non-expert users.

Dashboard automation or recommendation is another major area of research; Zeng \& Battle survey techniques for partially or fully automating the dashboard design process~\cite{zeng2024systematic}. 
For example, Multi-Vision~\cite{wu2022multivision} generates charts based on columns selected from a given dataset with a learning-to-rank approach. 
DMiner~\cite{lin2024dminer} analyzes over 800 dashboard examples for design rules to formalize the composition and interactions of individual views to recommend dashboards. 
Dashbot~\cite{deng_dashbot_2022} adopts deep reinforcement learning techniques to model human behavior for dashboard design, rewarding discovery of useful insights. 
These systems focus on the accumulation phase by training models to predict views that should appear together. 
Our work addresses a different need: given visualizations that creators have already accumulated, we infer what analytical themes they address and identify complementary views to fill gaps. 

Beyond understanding dashboard composition, several systems explore how to represent creator intent. 
Pandey~\etal{}~\cite{pandey_medley_2022} introduced Medley, a dashboard composition system, wherein users specify analytical intents via an input panel. 
Pandey~\etal also proposed a set of mappings to logically group dashboard views. 
Bolt~\cite{srinivasan_bolt_2023} extends Medley by mapping NL utterances to user objectives for dashboard generation.
While Medley and Bolt focus on recommending dashboard views as output, our system takes a dashboard as the \emph{input} and suggests a reorganized and enhanced set of mini-dashboards to communicate dashboard insights based on the identified analytic goals.  
We also refine the original intent templates such that they can be synthesized based on the data fields and types. 
In addition, we aggregate complementary insights and reorganize the dashboard to help guide interpretation of these analytical~goals.

After accumulating views, creators must arrange them into a coherent structure. 
D-Tour~\cite{dhanoa2024dtour} helps authors create interactive guided tours for onboarding using narrative templates to make dashboards more navigable. 
Qu and Hullman~\cite{qu2018keeping} proposed consistency guidelines to keep multiple views aligned, addressing a key challenge in multi-view coordination. Shi \etal{}~\cite{shi2023reverse} applied transformer-based models to predict visual similarity between views for layout optimization, focusing on perceptual grouping. 
These approaches ensure dashboards are visually coherent but do not address analytical coherence—whether views are grouped by the analytical questions they answer.

\vspace{-.2cm}
\subsection{Automated Insight and Multi-View Synthesis}

Insight discovery is often a key outcome for analyzing data through visualizations~\cite{battle2023what}.
For example, Voder~\cite{Srinivasan2019Voder} allows users to interact with automatically extracted data facts, and explore connections between data facts and visualizations. 
While early work in chart recommendation focused on mapping data to chart types and visual elements~\mbox{\cite{moritz2018formalizing,hu2019vizml,luo2018deepeye},} more recent research~\mbox{\cite{deng_dashbot_2022,shi_calliope_2021,wang_datashot_2020,ding2019quickinsights}} focuses on extracting patterns (\ie data facts) that can be represented by charts. 

Based on this idea, several systems have explored automating natural language authoring for multi-view visualizations, including dashboards, data stories, and data comics. 
These systems typically begin by mining meaningful patterns from raw datasets, then selecting and sequencing data-driven insights to present in natural language sentences alongside visualizations. 
For instance, DataShot~\cite{wang_datashot_2020} visualizes data facts as a fact sheet.
Calliope~\cite{shi_calliope_2021} sequences data facts into a meaningful story with logical transitions using a Monte Carlo tree search algorithm.  
ChartStory~\cite{zhao2023ChartStory} is the closest prior work to ours; it takes a collection of charts, clusters them by specification similarity, generates layouts based on data comic design patterns, and produces captions. However, ChartStory focuses on the visual and specification similarity rather than analytical goals, thus emphasizing effective comic-style transitions and layouts, whereas we focus on analytical structures and gaps.
Recent work has also explored the challenge of organizing and communicating analytical findings produced in computational notebooks. Notable~\cite{li2023notable} provides on-the-fly assistance for identifying and grouping insights during notebook analysis to support data storytelling, sharing our goal of surfacing and structuring implicit analytical patterns. A related thread focuses on transforming notebook analyses into presentation formats~\cite{wang2024outlinespark,wang2023slide4n,zheng2022telling}. 
These systems address the output end of the reflection process for communicating findings once they are identified, whereas our pipeline addresses the earlier challenge of inferring what analytical themes exist and which views support them. 
Closest to our optional LLM enhancement, InkSight~\cite{lin2024inksight} leverages LLMs to synthesize inferred facts from charts in computational notebooks into natural language annotations, demonstrating a complementary approach to our template-based extraction. 

While the aforementioned work concentrates on end-to-end generation of insight narratives, recent \mbox{work~\cite{li2024coinsight, aodeng2025inrectable, hoffswell2025representing}} gives users more control in exploring automated insights by leveraging a network representation of data facts, which better accommodates the non-linear nature of data exploration, allowing viewers to traverse connections in multiple directions.
Recent research has also shifted from template-based NL generation to approaches leveraging LLMs. 
InsightLens~\cite{weng2024insightlens} a multi-agent framework to generate insights. ChartThinker~\cite{liu2024chartthinker} uses chain-of-thought for chart summarization. Jupybara~\cite{wang2025jupybara} integrates LLMs into notebooks for analysis and storytelling. Leva~\cite{zhao2024leva} and LightVA~\cite{zhao2024lightva} apply agent-based approaches to onboarding and iterative task generation. These approaches demonstrate LLMs' flexibility for open-ended analysis.
Our work takes a complementary approach using rule-based, deterministic algorithms to prioritize verifiability and traceability. 
However, we view the integration of LLMs as an orthogonal opportunity rather than an alternative paradigm. Our pipeline's modular structure means individual components could be enhanced or replaced with LLM-based techniques. 
We explore LLMs as an opt-in enhancement layer at each pipeline stage, and describe the implementations in Section~\ref{sec:system}.

\begin{figure}[t]
    \centering
    \includegraphics[width=\columnwidth]{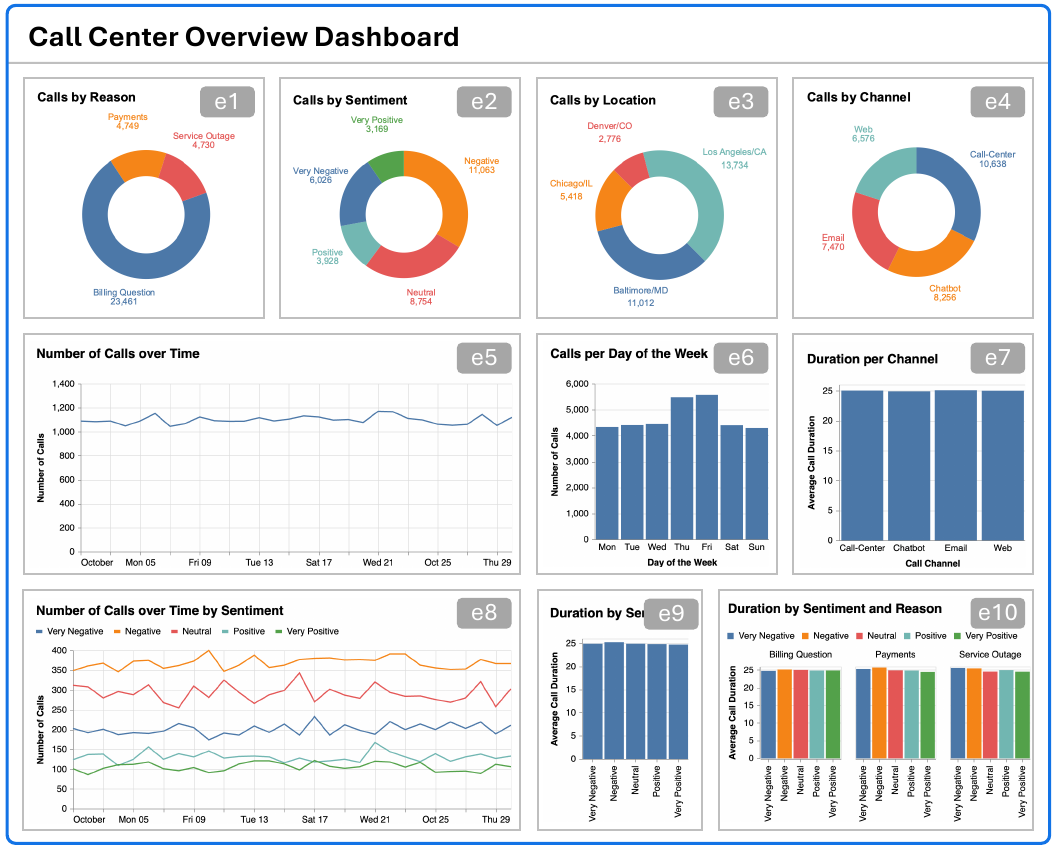}
    \caption{
    Ten charts and the preliminary layout for the call center dashboard described in the usage scenario (Section~\ref{sec:usagescenario}). For paper clarity, we label each \emph{existing} dashboard view in gray as \existingchart{1}, \existingchart{2}, etc. for easier reference in the text.
    }
    \label{fig:dashboard}
\end{figure}

\section{Usage Scenario}
\label{sec:usagescenario}

To illustrate how our pipeline can help creators refine their dashboards, we present the following fictional usage scenario: Quinn, a call center analyst, leverages our pipeline's automated analysis to iteratively refine her dashboard, highlighting how computational feedback can complement human expertise. 

\parHeading{Quinn creates a preliminary dashboard}
to understand the call center's performance and bottlenecks. 
Quinn first explores the data and creates ten views~(Fig.~\ref{fig:dashboard}) that investigate call volume, duration, and satisfaction across temporal, categorical, and operational dimensions. 
Quinn starts with univariate views to understand the basic distributions, such as call volume by reason~(\existingchart{1}), sentiment~(\existingchart{2}), location~(\existingchart{3}), and channel~(\existingchart{4}). After examining these breakdowns, Quinn adds temporal views showing call volume over time~(\existingchart{5}) and by day of week~(\existingchart{6}), as well as the duration by channel~(\existingchart{7}). 
Noticing that sentiment may exhibit temporal patterns, Quinn creates additional views exploring sentiment over time~(\existingchart{8}), as well as comparisons of sentiment by duration~(\existingchart{9}) and reason~(\existingchart{10}). 
Quinn's unfocused attention across multiple analytic themes (call volume, frequency, sentiment, duration, reason) yields a scattered collection of views in the initial dashboard. To organize her thoughts and refine the dashboard narrative,

\parHeading{Quinn now aims to assess the content and structure. }
A natural approach would be to arrange the dashboard according to her exploration process by showing the progression from basic distributions to temporal patterns and combined analyses. While this arrangement reflects Quinn's analytical journey, she wonders: \emph{Which analyses are key to summarizing call center performance?}
\emph{How do I want to structure the dashboard to highlight key findings?}
%
To help organize her thoughts and refine the dashboard narrative, Quinn runs our pipeline to get automated feedback, which detects three main themes: 
\begin{enumerate}
    \item \emph{Overall Volume Distribution}: summarize the overall call volume measure across different dimensions,
    \item \emph{Operational Efficiency+Satisfaction}: compare the average duration for the channel and call reason by sentiment,
    \item \emph{Temporal+Sentiment Patterns}: compare the call volume for both the sentiment and day of week over time. 
\end{enumerate}

Our pipeline also distills the insights for each theme by summarizing the low-level data facts, and organizes the goals, insights, and visualizations into a mini-dashboard. 
The result includes three mini-dashboards created based on the themes, each with the corresponding views and summarized insights.

\begin{figure}[!t]
    \centering
    \includegraphics[width=\columnwidth]{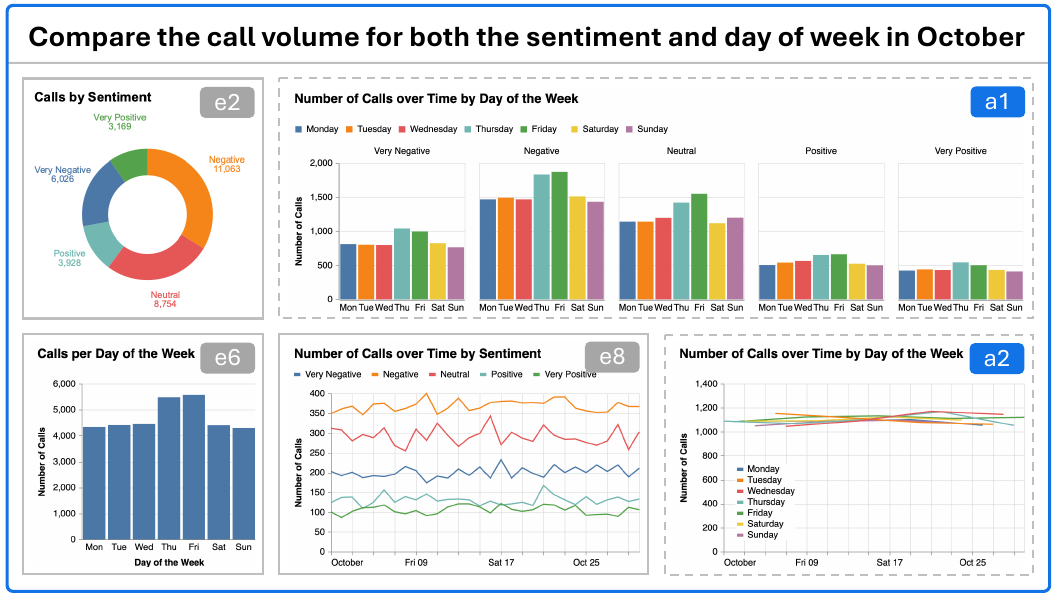}
    \caption{One of the three reorganized mini-dashboards produced by our pipeline. This mini-dashboard focuses on the third goal \emph{``compare the call volume for both the sentiment and day of week over time''} with two \emph{new} charts, \addedchart{1} and \addedchart{2}.
    }
    \label{fig:trend}
\end{figure}

\parHeading{Quinn evaluates the pipeline's recommendations. }
Quinn first reviews the third theme, \emph{``Temporal+Sentiment Patterns,''} which explores when negative sentiment spikes~(Fig.~\ref{fig:trend}). 
She notices that this theme has two new charts (Fig.~\ref{fig:trend} \addedchart{1} and \addedchart{2}) alongside the original charts related to the overall call volume and temporal attributes (Fig.~\ref{fig:dashboard}~\existingchart{2}, \existingchart{5},~and~\existingchart{8}). 
The first new chart (\addedchart{1}) combines the sentiment~(\existingchart{2}) and day of week~(\existingchart{5}) into a grouped bar chart, thus making it easier to see if certain days have a disproportionately negative sentiment. The second new chart (\addedchart{2}) shows temporal trends~(\existingchart{8}) for each day of the week~(\existingchart{5}), revealing if the patterns are temporally consistent.

The pipeline also provides aggregated insights, which state that \insightText{The call distribution is right-skewed, with Thursday and Friday having unusually high values. The pattern is consistent across all sentiments.}
Quinn notices that while both Thursday and Friday have higher absolute numbers across all sentiments, the ratio of negative sentiments appears higher for these two days compared to the others (Fig.~\ref{fig:trend}~\addedchart{1}). 
This observation prompts a hypothesis: \emph{perhaps the higher call volume on these days creates heavier workloads, leading to a worse customer experience.} While new data is needed to verify this hypothesis, the pipeline has surfaced a pattern worth investigating. 
Thus, Quinn decides to integrate \addedchart{1} into the main dashboard. 


\parHeading{Quinn refines the dashboard using her domain knowledge.}
While reviewing the second theme (\emph{``Operational Efficiency+ Satisfaction''}), Quinn ultimately disagrees with the pipeline's grouping, which combined duration and satisfaction together, despite them being separate concerns for the call center; based on her own expertise, call duration affects staffing and capacity planning whereas customer satisfaction affects service quality and retention. 
By reflecting on the pipeline's grouping, Quinn recognizes an opportunity to deepen the analysis; Quinn decides to split the theme into two separate mini-dashboards that better reflect the call center context:
(2a)~\emph{``Duration+Capacity Patterns''}: summarize the call duration across both channels and clients, and 
(2b)~\emph{``Satisfaction~Analysis''}: summarize the customer satisfaction and sentiment, and explore new charts.

\parHeading{Quinn prepares to present to key stakeholders.}
The pipeline output serves multiple purposes in Quinn's analytic workflow. The grouped themes help Quinn articulate what the dashboard investigates, the enriched views fill analytical gaps, and finally the natural language insights can help provide a starting point for communicating essential findings to key stakeholders. For example, one insight from the \emph{``Duration\,+\,Capacity Patterns''} theme explains that \insightText{Billing Question shows the smallest variation of average call duration across channels, while others range from 0 to $\sim$25 minutes.}
Quinn is thrilled that the pipeline has helped distill the key insights for each topic in natural language, which reduces some of the time and effort required to create a detailed report on the findings. 

\parHeading{Usage Scenario Reflections.} During the refinement process, the pipeline provides a computational perspective while Quinn provides contextual judgment. In particular, the pipeline identified key structural patterns and surfaced data relationships, but Quinn's domain expertise was essential for recognizing that the initial grouping conflated operationally distinct concerns. This combination of automated analysis and human refinement exemplifies how our pipeline can be used as a design reflection tool that augments rather than replaces analyst expertise.

\section{Dashboard Remixing Pipeline}
\label{sec:system}

\pipeline
We leverage a multi-stage pipeline to transform exploratory dashboards into reorganized mini-dashboards, each with insights summarizing patterns relevant to the inferred analytical goals. 
The process begins by analyzing the input dashboard specification (Fig.~\ref{fig:pipeline}\,\textcircled{\footnotesize A}), which includes the chart designs and underlying data. 
We enumerate potential analytical goals by mapping dashboard attributes to goal templates~(Fig.~\ref{fig:pipeline}\,\textcircled{\footnotesize 1}). 
Next, we employ mixed integer linear programming (MILP) to optimally select the most relevant goals that best explain the dashboard while minimizing redundancy (Fig.~\ref{fig:pipeline}\,\textcircled{\footnotesize 2}). 
For each selected goal, we generate data insights from both the existing and newly synthesized charts, scoring them based on their relevance to the analytical objective (Fig.~\ref{fig:pipeline}\,\textcircled{\footnotesize 3}).
Finally, we aggregate the data facts connecting the dashboard views, to create consolidated insights that weave scattered evidence into analytical threads with specific objectives (Fig.~\ref{fig:pipeline}\,\textcircled{\footnotesize 4}). 
The pipeline produces a webpage as the final output~(Fig.~\ref{fig:pipeline}\,\textcircled{\footnotesize D}), where users can quickly scan the recommended organization, views, and insights for useful ideas to refine the design.  

\vspace{-8px}
\subsection{Input Dashboard Specification} \label{sec:spec}

In this section, we describe how we model dashboards within our pipeline.
A dashboard $D$ is a collection of chart views, where each view $e_i$ is specified by a \textit{data scope} and its \textit{charting domain specific language (DSL)}. 
We follow the definition of data scope from Ding~\etal~\cite{ding2019quickinsights}, and visual mappings inspired by Vega-Lite~\cite{satyanarayan2017vegalite}. 
Formally, a view $e_i$ (where $e$ stands for ``existing views'') is defined~as:
\begin{align*}
    e_i &:= <data\ scope, charting\ DSL> \\
    data\ scope &:= <subspace, breakdown, measure, agg> \\
    agg &:= count|mean|sum \\
    charting\ DSL &:= f(data\ scope, c_x, c_y, c_{color}, c_{facet}) \\
    f &:= Bar|Donut|Line|Scatter
\end{align*}

where \textit{subspace} is defined by a set of filters $\phi_i$ applied on the dimensions of the original data; 
a \textit{breakdown} dimension is a categorical or temporal column where the group-by operator is performed to divide the subspace;
a \textit{measure} is a quantitative column that an aggregate $agg$ can be calculated on.  
The charting DSL takes a data scope as the input table, and maps its attributes $c_x$ and $c_y$ to the x-axis and y-axis respectively;
$c_{color}$ and $c_{facet}$ are specified only for multi-view or multi-layered charts (\eg multi-line charts, small multiples, \etc). 
For example, a donut chart is essentially a stacked bar chart with a single bar where the numeric values use angular extents.
Thus, the donut chart \existingchart{1} from the example dashboard in Fig.~\ref{fig:dashboard} can be specified as $Donut(data\ scope, c_y=calls, c_{color}=reason)$.
Following this dashboard specification framing, our pipeline~is extensible to more chart types such as bubble or density charts. 


\subsection{Step \textcircled{\footnotesize 1}: Analytic Goal Enumeration} 
\label{sec:system:goal-enumeration}

{Given an input dashboard, we infer the analytical goals its existing views collectively express to a viewer, i.e., the goals the dashboard supports as constructed, rather than the implicit intent known only to its creator. }
We extend existing work on defining creator intents to allow for simultaneous assessment of different goal types;
we distill five high-level design intents inspired by prior work on dashboard automation~\cite{pandey_medley_2022, srinivasan_bolt_2023}, and formally define them as goal templates~(Section~\ref{sec:goal_temp}).
However, a template is a generic pattern for how a creator might design a dashboard, and must be populated to yield a tangible goal (i.e., an \emph{instance} of the template).

A goal instance is essentially a program specifying a group of charts. 
To generate goal instances, the template is populated by filling the holes with a valid combination of data attributes and transformations based on its definition.
Each goal template is associated with a set of specific view types that can be used to achieve a goal~\cite{pandey_medley_2022}.
Leveraging this approach, each generated goal instance can be used to instantiate a set of pre-defined view specifications. 
For example, for the sample dashboard in Fig.~\ref{fig:pipeline}, the instantiated goal $g_1$ (\texttt{SummarizeMeasure}) corresponds to seven charts shown in mini-dashboard $M_1$. 

However, if the input dataset has many attributes, there~may be a prohibitively large set of possible attribute combinations for populating our goal templates, producing an intractable search space of goal instances. 
To mitigate this problem during goal enumeration, we generate all possible goal instances that instantiate at least one of the existing dashboard views to reduce the search space for the next phase. 
At the same time, we annotate whether or not the views of each goal instance are existing or new. For example, the two line charts are marked as existing and the two bar charts as new for $M_2$ in Fig.~\ref{fig:pipeline}\,\textcircled{\footnotesize C}.

\parHeading{Goal templates:}\label{sec:goal_temp} 
Our goal templates take a data table~$T$ as input and output a collection of views ${\{e_i, ..., e_j\}}$. We currently support five types of goal instances:
(1)~summarize a single measure, (2)~compare two measures, (3)~summarize how a single measure changes over time, (4)~compare categories in a single dimension, and (5)~compare two dimensions.

\parHeadingNoSpace{(1) Summarize a single measure:} 
This template includes a set of views that summarize a quantitative measure $q$ across more than two categorical, temporal, and geographical attributes~$\overline{a}$.
For example, as shown in Fig.~\ref{fig:mini} (left), a viewer may want to understand 
the number of movies per rating~($e_1$), year~($a_1$), and actor~($e_7$).
When the aggregation function $\alpha = COUNT(*)$ is applied as the measure $q$, the goal can be interpreted as displaying univariate summaries profiling the data attributes. 
When the attributes $\overline{a}$ include columns with a large number of unique values, this goal can also be interpreted as observing the top N units of such a column. 
{For example, Fig.~\ref{fig:mini} \existingchart{7} shows the top twenty actors with the largest number of movies.}
    
\parHeadingNoSpace{(2) Compare two measures:} 
This template enumerates a set of views that compare between two quantitative measures $q_1$ and $q_2$, across more than one categorical, temporal, and geographical attributes $\overline{c}$, to understand their relationship. 
The template generates a scatter plot with $q_1$ and $q_2$ on each axis, directly encoding their relationship. To complement these scatter views, juxtaposed bar or line charts show each measure separately, supporting comparison across discrete categories or over time. 
For example, we may want to understand correlations between call durations and total calls per day across call center locations, or understand potential relationships between IMDB ratings and total votes{ (e.g., Fig.~\ref{fig:mini}, right)}.

\parHeadingNoSpace{(3) Summarize change over time: } 
This template displays the value of a quantitative measure $q$ over the temporal values of a different time unit (\eg day of week) $timeunit(T, unit)$, and across other attributes $\overline{a}$.
In particular, this goal template supports trend analysis, seasonality detection, and temporal pattern recognition across other categorical dimensions, helping users understand both how and when changes occur in their data. 
For example, how has the call volume changed over the last year across different regions?

\parHeadingNoSpace{(4) Compare categories in a single dimension:} 
This template includes views that highlight the difference between the quantitative value $q$ of categories from a categorical attribute $c$ over more than two other attributes $\overline{a}$. 
In other words, this goal examines how quantitative values vary across categories like channels, sentiments, or regions, enabling comparative analysis and identification of high or low-performing categories across multiple contexts. 
For example, which call centers might be under-performing compared to all regions? Which film genres tend to have higher budgets?

\parHeadingNoSpace{(5) Compare two dimensions:} 
This template shows the views highlighting the difference between the quantitative value $q$ of the categories from two categorical attributes $c_1$ and $c_2$ over another attribute $a$.
This goal thus examines how quantitative measures vary across combinations of dimensions (e.g., channel types and call reasons), revealing interaction effects and identifying optimal or problematic combinations. 
For example, the mini-dashboard from our usage scenario in Fig.~\ref{fig:trend} examines the relationship between sentiment and day of the week in terms of the call volume.


Each goal template defines a subspace of potential goal instances of interest, which imposes a beneficial structure that prevents the system from generating nonsensical goals.
To this end, the goal enumeration step generates goal instances that are syntactically correct according to the given template. 
Specifically, for each template, our search algorithm selects every combination of parameters that match the types assigned within the template, as described above. 
{By formalizing these common analytical intents, our system can systematically identify and extend meaningful analytical narratives that the dashboard's views already express, giving creators a computational rendering of their design's analytical structure to compare against their intended design goals.}

We acknowledge that these goal templates are not exhaustive and the mapping could be more flexible.
However, these templates represent common practice and knowledge surveyed by the prior literature~\cite{pandey_medley_2022,srinivasan_bolt_2023}. 
Further, our approach can be extended with customized templates in the future. 

\subsection{Step \textcircled{\footnotesize 2}: Optimal Goal Selection}
\label{sec:system:step2}

While the system must reason about all relevant intents for a given dashboard, it must also avoid overwhelming the user with too much information. 
Thus, the system must eventually pare these intents down to a much smaller set that should be used to generate recommendations.
In the second step of our pipeline, we infer the most relevant intents to effectively represent the dashboard (see Fig.~\ref{fig:pipeline}, pipeline \textcircled{\footnotesize B} step \textcircled{\footnotesize 2}).  
This selection problem can be formalized as a weighted set cover problem and solved using mixed integer linear programming (MILP).
Our optimization approach aims to identify the set of goal instances that best explain the existing dashboard views while minimizing redundancy.

We used Gurobi~\cite{gurobi} as the solver for our MILP formulation.
The solver returns the optimal set of goals for the input dashboard views, which guides the analytical structure and assists in generating insight bullet points relevant to each goal.


\parHeadingNoSpace{Input:} a dashboard consisting of $n$ existing views $D = \{e_1, e_2, ..., e_n\}$. 
Each view represents a visualization with specific data attributes and encodings as defined in Section~\ref{sec:spec}. 

\parHeadingNoSpace{Additional views:} let $A= \{a_1, a_2, ..., a_m\}$ be the newly generated views that do not exist in the dashboard input. 

\parHeadingNoSpace{Goals:} a set of goal instances $G$, where each goal $g = \{D_g \cup A_g : D_g \subseteq D, A_g \subseteq A\}$ requires a set of views.
Each goal represents a potential analytical intent of the dashboard, such as ``summarize change in call volume over this year.''

\parHeadingNoSpace{Charts:} let $C = D \cup A$ be the union of all existing dashboard views and additional views.
This union represents the complete pool of visualizations available for selection.

\vspace{8px}\noindent The main optimization challenge is to select a subset of goals that maximizes coverage of existing dashboard views in $D$ while minimizing the introduction of new views from $A$, and set a threshold for the maximum number of goals necessary to explain the dashboard.

\parHeading{Define the Objective Function:} 
Our optimization function balances three competing objectives when selecting the goals: (1) maximizing dashboard coverage, (2) minimizing additional views, and (3) minimizing the number of goals.

\parHeadingNoSpace{(1) Maximizing dashboard coverage:}
We prioritize the goals covering as many existing views as possible, since they are the most likely to represent the dashboard's existing analytical structure.
This objective is formalized as maximizing $ \sum_{i \in C} y_i \cdot \1{i \in D}$, 
where $y_i$ is a binary variable indicating whether view $i$ is covered by any selected goal, 
and $\1{i \in D}$ is an indicator function that equals 1 if view $i$ is from the original.

\parHeadingNoSpace{(2) Minimizing additional views:}
To prevent overwhelming users with too many new visualizations, we penalize their occurrence. 
This objective is represented as follows, wherein the goal is to minimize $ \sum_{i \in C} y_i \cdot (1 - \1{i \in D})$.

\parHeadingNoSpace{(3) Minimizing the number of goals:}
To ensure a concise narrative structure, we aim to explain the dashboard with as few analytical goals as possible. 
This objective is captured by minimizing $\beta \sum_{g \in G} x_g$, where $x_g$ is a binary variable indicating whether goal $g$ is selected, and $\beta$ is a constant. 
We then define the combined objective function as follows:
\vspace{-1px}
$$\text{\footnotesize{Maximize}} ( \sum_{i \in C} y_i \cdot \1{i \in D}  - \alpha  \sum_{i \in C} y_i \cdot (1 - \1{i \in D}) - \beta \sum_{g \in G} x_g)
$$
Weights $\alpha$ and $\beta$ are hyperparameters that can be tuned based on the usage scenario and user preferences. 
Higher values of $\alpha$ discourage the introduction of new views, while higher values of $\beta$ result in simpler narrative structures with fewer groups.

\parHeading{Define the Constraints:} 
We impose the following constraints on the overall objective to ensure a reasonable solution:

\parHeadingNoSpace{(1) Selected views are supported by the goals:}
For each view in the grouping results, the constraint $\sum_{g \in G} \cdot \1{i \in g} x_g \geq y_i \quad \forall i \in C$ 
creates a logical dependency between the view selection variable $y_i$ and the goal selection variable $g_i$, 
which guarantees that the view is covered by the selected goals.

\parHeadingNoSpace{(2) At least one goal (and at most k goals) must be selected:}
At least one goal should be selected to prevent empty results, which can be expressed as $\sum_{g \in G} x_g \geq 1$. 
For complex dashboards and datasets, we may also introduce an upper bound constraint to limit the total number of selected goals $k$ to prevent cognitive overload.
For practical reasons, we set an upper bound of $k=5$, and leave it to future work to fine-tune optimization parameters across scenarios.


\subsection{Step \textcircled{\footnotesize 3}: Data Fact Extraction}
\label{sec:extract}



\noindent Our system adapts established techniques for data fact extraction~\cite{deng_dashbot_2022,ding2019quickinsights} to identify statistically significant patterns and relationships within the visualizations. 
Our goal selection process in Step \textcircled{\footnotesize 2} yields a set of goals $G$, where each goal $g \in G$ represents a set of charts, some of which may already be present in the dashboard.
For each set of charts, we employ a fact generation process that produces a set of candidate insights.
Our fact generation approach distinguishes between two fundamental types of insights from the literature: 
atomic facts derived from individual series and cross-series facts that emerge from comparisons across multiple data series.
Similar to prior work~\cite{ding2019quickinsights,wang_datashot_2020,shi_calliope_2021}, we specify a data fact $F_i$ as:
\begin{align*}
    F_i^j &:= <e_i, type, description, properties> 
\end{align*}
where $e_i$ is the view identifier, \textit{type} is the data fact type that is constrained by the chart type and data scope of $e_i$, \textit{description} is the natural language description generated from a template for each type, and \textit{properties} are the supporting values that fill the template. 
For example, the call center overview dashboard in our usage scenario (Fig.~\ref{fig:dashboard}) generates an atomic data fact of type \textit{extreme} for chart \existingchart{5}, which states that 
\insightText{in the Bar chart of calls by day, Friday~has the highest value at 5570.0.}
The properties for this data fact are specified as \{``subtype'':~``maximum'', ``item'':~``Friday'', ``value'':~5570\}. 
For visualizations containing multiple series (e.g., grouped bar charts, multi-line charts), we extract cross-series facts, enabling the discovery of complex patterns and relationships among multiple data series. 
For example, \insightText{Call-Center shows the greatest variation between Neutral and other series, with values ranging from 24.3 to 25.4} compares the time-series of call amounts among different channels. 

Each data fact receives an importance score based on its statistical significance and its priority to the chart type. 
Generally, pattern-related facts  (\eg trends) are preferred over point-related facts (\eg extrema).
We employ different extractors based on common chart types since they reveal different types of insights. 
These patterns often generalize across chart types because the underlying data queries are similar~\cite{battle2023what, purich2025adaptive}. 
Thus, we focus on four chart types for their coverage of aggregate, proportional, and temporal factors:

\begin{itemize}
    \item \textbf{Bar Charts:} Extract extremes, outliers, distribution patterns such as skewness and evenness, and patterns like monotonic trends or segmented comparisons.
    \item \textbf{Donut Charts:} Identify proportions, dominant segments, and concentration patterns where multiple categories collectively account for significant percentages.
    \item \textbf{Line Charts:} Detect temporal trends, significant changes between consecutive points, seasonality, and convergence or divergence patterns in multiple series.
    \item \textbf{Scatter Plots:} Identify directional relationships and correlations between two quantitative measures, including positive or negative associations, clustering of categories, and outlying entities that deviate from the overall pattern.

\end{itemize}

To extend the pipeline to more chart types, the mapping from chart type to extractor needs to be specified, although the individual extractors could be re-used based on the data~types. 

\parHeading{Optional LLM Enhancements.} 
The default extraction dispatch is deterministic: given a chart type, a fixed set of extractors is applied regardless of analytical context. We additionally support two optional LLM-based modes. In the \textit{agentic extraction mode}, chart metadata and the enclosing analytical goal are sent to an LLM, which uses tool calling to select which extraction functions to invoke for each chart. 
The extraction functions themselves remain unchanged so that the LLM acts as a goal-aware selector that prioritizes fact types relevant to the analytical goal 
without performing the underlying computation. Additionally, after extracting structured facts, an LLM can optionally rewrite the human-readable description field for improved fluency in the \textit{fact description mode}. 
The agentic mode affects the downstream aggregation step while the description mode does not, since it only consumes the structured fact objects without the textual descriptions. 
\subsection{Step \textcircled{\footnotesize 4}: Insight Aggregation}
\label{sec:aggregation}

A significant challenge in generating structured narratives from dashboards comes from transforming individual data facts into coherent stories~\cite{hoffswell2025representing}.
Traditional data storytelling systems focus on building sequential narrative transitions
 via interpolated analysis paths or visualization similarity~\cite{sun2022erato, kim2017graphscape,zhao2023ChartStory}.
However, dashboards are exploratory by nature and dashboard views are often designed to cover different data slices in parallel. 
This flexibility poses a unique challenge: since the views do not necessarily follow a clear narrative order, neither do the insights extracted from them. 
Instead, insights can be seen as nodes on a dense network with implicit connections that must be synthesized~\cite{hoffswell2025representing}.
We thus contribute a multi-level insight aggregation framework that addresses this challenge by hierarchically consolidating related data insights.

By constructing a \textbf{semantic network} that explicitly represents relationships between data facts, we can identify meaningful connections across disparate dashboard views and aggregate them into cohesive narrative units. 
Starting from basic chart-specific observations, our pipeline \textbf{aggregates data facts} into more complex insights that capture entity-specific patterns and cross-entity relationships, where specific observations are synthesized into broader utterances in alignment with inferred analysis goals (Section \ref{sec:system:step2}).

\parHeading{Semantic Network Construction:}
We construct a semantic graph that serves as the foundation for fact pattern detection and aggregation by encoding how facts relate to each other. 
The network is structured as a directed graph where nodes represent data fact objects and edges represent relationships between individual data facts based on their data scope and the item to which it refers. 
We build the network while extracting facts from the charts for each goal by iteratively adding the new data fact and assessing it against the existing nodes to establish relationships as the edges. 
We support three primary types of relationships between data facts:

\parHeadingNoSpace{(1) Contrasting} relationships connect facts associated with different categories of the same dimension and data scope. 
For instance, a maximum fact (\eg \insightText{email having the highest customer satisfaction}) and a minimum fact (\eg \insightText{phone having the lowest customer satisfaction}) that describe the same measure (satisfaction) across the same dimension (channel) naturally form a contrasting pair. 

\parHeadingNoSpace{(2) Drill Down} relationships represent generalization and elaboration between data facts based on either data scope (one is aggregated from the other) or fact type (point-fact versus shape-fact under the same data scope), where the inner data fact provides more context for the outer data fact. 
These relationships can also be identified through known pairs: trends drill down to change points, and distributions drill down to outliers. 
For instance, a trend data fact about \insightText{customer satisfaction declining over time} provides context for a change data fact about \insightText{a significant drop in satisfaction last month.} 

\parHeadingNoSpace{(3) Generally Related} is a default relationship between two data facts that are associated by types or entities including dimensions, measures, and categories. 
These relationships connect facts analyzing the same entity from different dimensional perspectives and facts analyzing related measures (\eg \insightText{Email takes the longest average duration} and \insightText{Email has the worst average satisfaction score}). 

\parHeading{Summary:} These relationships are our foundational fact aggregation patterns, which can be surfaced by well-defined rules. Although future work might incorporate
domain knowledge to provide semantic context for more flexible and dynamic relationships~(e.g.,~\cite{battle2023what}). We will discuss this in the expert review and discussion.

\parHeading{Insight Aggregation and Ranking:}
Building on the semantic network of data facts, we compose higher-level insight patterns by detecting subgraph structures that correspond to meaningful analytical narratives. 
These patterns emerge from the logical combination of semantically related facts, creating aggregated insights that provide richer analytical value than isolated~facts.


\begin{figure*}[t]
    \centering
    \includegraphics[width=.9\textwidth]{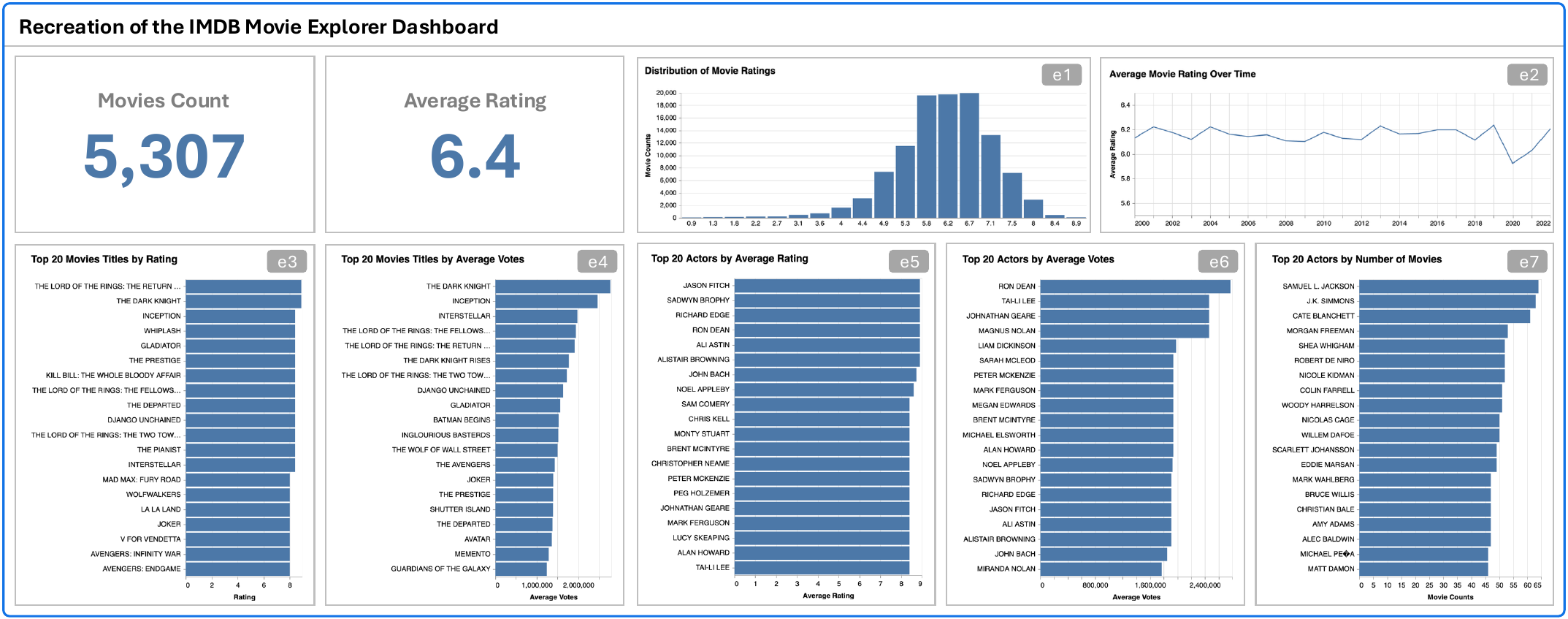}
    \caption{Recreation of a movie dashboard~\cite{movieDashboard} (D1) from Tableau Public for the use case in Section~\ref{sec:usecase}. 
    We label each original dashboard view in gray as \existingchart{1} through \existingchart{7}:
    a histogram showing the rating distribution~(\existingchart{1}), a line chart tracking average ratings~(\existingchart{2}), two bar charts for the top-rated~(\existingchart{3}) and most-voted movies~(\existingchart{4}), and three bar charts for the top actors with the highest average ratings~(\existingchart{5}), the {average} votes~(\existingchart{6}), and the most movies~(\existingchart{7}). The two KPI tiles are included to replicate the original dashboard but are not processed by the pipeline, as they do not correspond to any goal template.
    }
    \label{fig:movies}
\end{figure*}

We define \textit{insight patterns} as compositions of multiple facts connected through specific relationship structures. 
Each pattern represents a distinct analytical narrative template that can be instantiated across different data scopes and entities. 
For example, a \textit{hierarchy pattern} emerges for a subgraph containing: (1) a fact representing overall leadership,  (2) additional facts of the same type, and (3) drill-down relationships between the overall leader and context-specific leaders. This composition reveals the analytical insight that ``Entity X leads overall, but Entity Y dominates in specific contexts.''
A \textit{consistency pattern} forms when multiple facts of the same semantic direction (all outperform or all underperform) are connected through the contrasting relationship type. This pattern composition identifies entities that maintain reliable performance across various analytical dimensions. Similarly, a \textit{contradiction pattern} associates facts with opposite semantic directions. 


We implement pattern detection through subgraph matching on the semantic network. For each pattern, we define constraints on fact types and the relationship that connects these data facts.
The algorithm traverses the network to identify subgraphs matching these structural requirements. 
Each detected pattern generates an NL narrative using a template 
populated with entity names, values, and directional relationships drawn directly from the matched subgraph.

Not all detected patterns are equally relevant for every analytical goal. 
We implement a goal-specific scoring system that weights patterns and facts based on their analytical utility for different goals. 
Each analytical goal type (compare dimension, summarize measure, time analysis, etc.) maintains weights for different pattern and fact types based on their capacity to answer the goal's core analytical questions.
For instance, hierarchy patterns receive high relevance scores for ``compare dimension'' goals because they directly address the question \emph{``which category performs best overall versus contextually?''} However, the same hierarchy pattern receives lower scores for the ``summarize change over time'' goal where temporal trends are more valuable than static performance rankings. 
Our current implementation uses several hyperparameters such as  confidence thresholds for fact extraction and relationship weights that were empirically tuned on development examples. 
Future work could learn these parameters from user feedback.

\parHeading{Optional LLM Narrative Enhancement.}
We additionally provide optional LLM-based narrative enhancement at two levels. Approach A post-processes rule-based output: after the semantic network detects patterns and composes groups, an LLM rewrites each group narrative given the detected pattern type and its constituent fact descriptions, improving fluency while retaining the rule-derived structure. The detected pattern type (hierarchy, consistency, or contradiction) is determined by rule, and the LLM only improves how the narrative is expressed. Approach B bypasses rule-based aggregation entirely: raw facts for each goal are sent to the LLM, which groups them into insight clusters and writes narratives directly, without the semantic network intermediate step. 
Both approaches introduce variability across models and prompt wordings, which is problematic for long-term use that requires continuous assessment and tuning. 

\subsection{Dashboard Remixing Output}
Finally, our pipeline generates a webpage that reorganizes the views and consolidates data facts based on the inferred goals. For example, Fig.~\ref{fig:pipeline}\,\textcircled{\footnotesize A} shows a recreation of a hospital ER dashboard from Power BI~\cite{erDashboard} and Fig.~\ref{fig:pipeline}\,\textcircled{\footnotesize D} shows the resulting webpage from running our pipeline.
Dashboard views and enriched views are grouped by goals, while insights are displayed in a side panel. 
Users can interact with insights through hover actions, which highlight the contributing charts. 
Clicking on the insights expands the insight panel to show the supporting data facts. 
Clicking on individual charts reveals the chart-level facts extracted from that specific view. 
Rather than directly producing a final dashboard, our pipeline surfaces the dashboard's expressed analytical structure, which creators can compare against and refine toward their original intents.

The pipeline is a standalone Python application where the user provides a dashboard specification described in Section~\ref{sec:spec} together with the underlying dataset. 
It supports two modes: a batch mode that runs the full pipeline and writes a self-contained HTML output, and an interactive server mode that serves the output in a browser, enables revised selection via chart bookmarking, and triggers re-runs to produce a new set of mini-dashboards reflecting the creator's revised focus. 
For the dashboards in our evaluation, the default rule-based pipeline completes in under 3 seconds for each iteration on a standard laptop. Runtime scales primarily with the number of views, which remains tractable for typical dashboards. The optional LLM-enhanced mode requires approximately 1 minute, dominated by LLM API (GPT-4.5) latency.

\section{Evaluation}

We evaluate our pipeline through two complementary approaches: 
(1) a discussion of four example use cases adapting dashboards collected from Tableau Public and the Power BI gallery demonstrating the pipeline's versatility across domains and levels of complexity, and (2) feedback from six dashboard experts examining the usability. 
Our expert interviews are guided by two research questions: 

\parHeading{RQ1:} How do the result of the proposed dashboard remixing pipeline compare to the original dashboard for identifying the analytical goals and summarizing important takeaways?

\parHeading{RQ2:} How do dashboard experts perceive the overall quality of the generated analytic goal groups and insights?

\subsection{Dashboard Use Cases}
\label{sec:usecase}
We selected four real-world dashboards from different domains including customer service, healthcare, operations, and movies. 
The number of views for each dashboard ranges from 7 to 11 on a single tab. Their source data contains categorical, temporal, and quantitative data types. 
We discussed the call center overview dashboard in Section~\ref{sec:usagescenario}; in this section, we discuss two
new dashboards (D1 and D3).
{Images for the dashboards and pipeline outputs are included in the supplemental material, along with a discussion of the HR use case (D2).}

\newcommand{\movieDashboard}{D1}
\newcommand{\HRDashboard}{D2}
\newcommand{\hospitalDashboard}{D3}


\begin{figure*}[t]
    \centering
    \includegraphics[width=.9\textwidth]{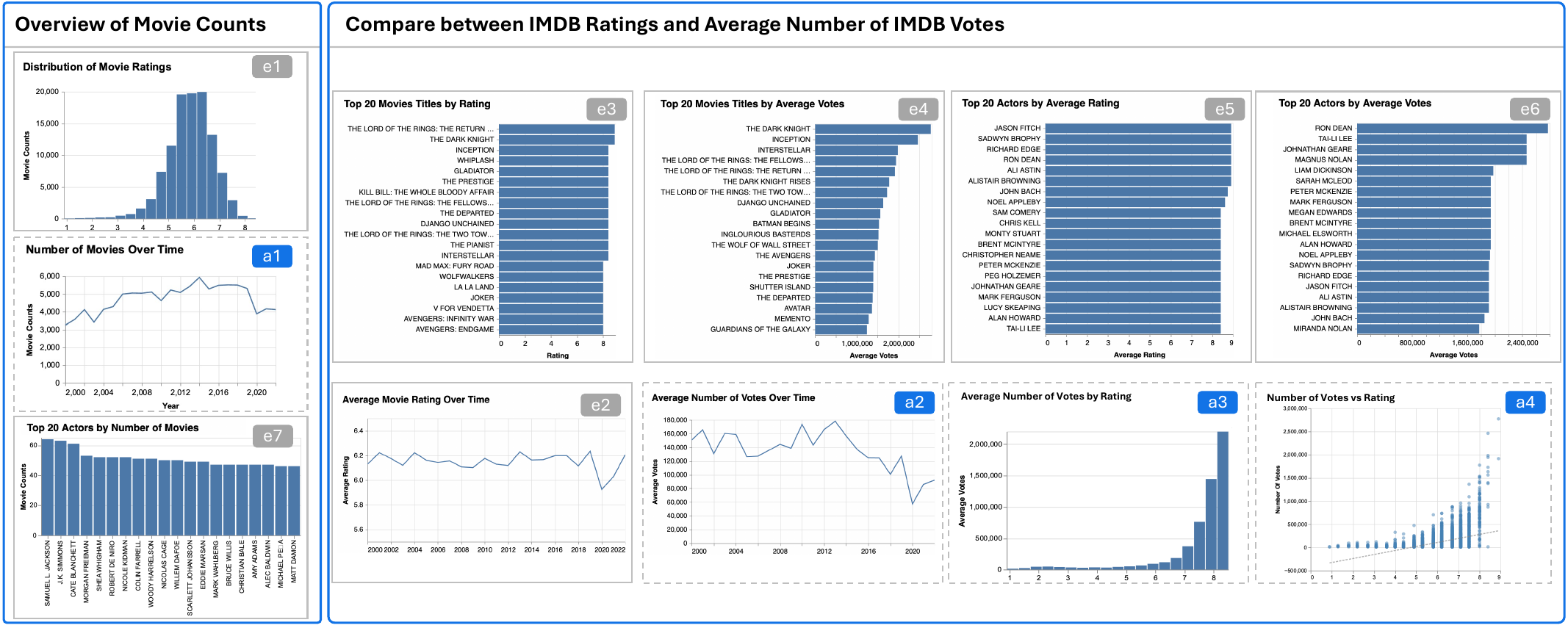}
    \caption{Two mini-dashboards are recommended based on the input dashboard in Fig.~\ref{fig:movies}, to address two analytic goals identified by our pipeline: overview of movie counts (left) and comparison of ratings and votes (right). 
    Four new charts are added to enrich these analytic goals: a line chart showing the number of movies per year (\addedchart{1}) extends the summary analysis for the overview charts exploring the number of movies by rating (\existingchart{1}) and actor/actress (\existingchart{7}); a new line chart of average votes over time (\addedchart{2}) complements the one showing average ratings over time (\existingchart{2}); finally, a new histogram and a scatter plot show the distribution relationship between ratings and average votes (\addedchart{3} and \addedchart{4}), making the comparison of metrics more explicit.}
    \label{fig:mini}
\end{figure*}

\parHeading{(D1)~IMDB Movie Explorer}~\cite{movieDashboard}
contains movie records from 2000 to 2022, with attributes including movie titles, actors/actresses, release years, IMDB ratings, and vote counts.
It incorporates seven visualizations (labeled \existingchart{1}-\existingchart{7} in Fig.~\ref{fig:movies}). 
The top charts show the rating distribution and average rating by year. 
The bottom views rank the movies and actors by rating, average number of votes, and total number of movies the actor starred in. 
The input dashboard is a typical exploratory dashboard organized into four panels, each containing a single chart or table. 
While the overview of movie counts is distributed across the original dashboard (\existingchart{1}~and~\existingchart{7}), it is distilled and augmented with a new line chart showing the number of movies each year~(\addedchart{1}) in our first mini-dashboard (Fig.~\ref{fig:mini}, left). 

Furthermore, the comparison between ratings and votes, while clearly of interest to the dashboard creator based on their repeated presence and proximity in the original dashboard, is generally implicit and requires users to draw their own conclusions. 
Our second mini-dashboard (Fig.~\ref{fig:mini}, right) goes a step further by providing direct comparisons.
To complement the view showing the average rating over time~(\existingchart{2}), a new~line chart shows the average votes over time (\addedchart{2}); 
a new histogram showing the distribution of average votes across ratings directly demonstrates the relationship between the two~(\addedchart{3}). 

Finally, aggregating the data facts makes it easier to reflect on interesting patterns and outliers in the data. 
For example, one insight reveals that the average votes and the average rating show a moderate positive correlation over time: \insightText{Average Votes and Average Rating show a moderate positive correlation (r=0.50), indicating they tend to move together. This suggests complementary dynamics between these measures.} This result can be difficult to discern from \existingchart{2} and \addedchart{2} alone, due to the incompatible scales (0-10 ratings versus thousands of votes). 
The aggregated insights also capture the pattern in \addedchart{3} demonstrating that movies with higher ratings tend to have larger average number of votes: \insightText{The Bar chart shows a consistent increasing pattern in Average Number of Votes across Ratings, indicating a positive relationship between their~values.} 


\parHeading{(D3)~Hospital Emergency Room Dashboard}~\cite{erDashboard}
presents operational analytics for a hospital ER. 
The dashboard records 1,338 fictional patients over 18 months, tracking the wait time, satisfaction score, and admission status. 
The input dashboard in Fig.~\ref{fig:pipeline}\,\textcircled{\footnotesize A} has eleven visualizations, grouped into three mini-dashboards by \emph{demographic analysis} (age group, gender, and racial distribution), \emph{operational statistics} (department referral, admission status, and patient visiting day and time), and \emph{temporal trends} (admission status, average wait time and average satisfaction score over time). 
This dashboard organizes views based on the semantic domain, which conforms naturally to how analysts apply domain knowledge during the reasoning process. 
Patient demographics are clustered together to analyze the patient population, while operational metrics help identify bottlenecks to improve efficiency, and monitoring KPI changes over time provides opportunities to detect drastic changes.

In contrast, our pipeline groups the views into four specific goals. 
The first mini-dashboard (Fig.~\ref{fig:pipeline}\,\textcircled{\footnotesize C}, M1) summarizes the patient volume by selecting the trend line and the bar charts of various dimensions. In addition, the pipeline generates new bar charts showing the gender distribution and race distribution. 

The second mini-dashboard (Fig.~\ref{fig:pipeline}\,\textcircled{\footnotesize C}, M2) compares the average wait time and average satisfaction over time and across dimensions. 
As shown in the pipeline output from Fig.~\ref{fig:pipeline}\,\textcircled{\footnotesize D}, multiple charts reveal that categories with higher wait times have lower scores, which is also captured by the first goal insight: \insightText{categories with lower average wait times show higher average satisfaction scores.} 
However, the correlation fact examining individual patient-level relationships indicates that \insightText{wait time and score show no linear correlation \mbox{(r=-0.02).}} 
One explanation for these seemingly contradictory patterns is that group-level trends may not translate to individual correlations due to Simpson's paradox and within-group variance. 

The third mini-dashboard (Fig.~\ref{fig:pipeline}\,\textcircled{\footnotesize C}, M3) compares the patient volume across the day of week they visited the ER and whether they were admitted. 
The last group (Fig.~\ref{fig:pipeline}\,\textcircled{\footnotesize C},~M4) compares the volume across patient races and age ranges, similar to the original ``Patient Demographic'' panel in Fig.~\ref{fig:pipeline}\,\textcircled{\footnotesize A}. 

In contrast, our reorganized mini-dashboards reveal a similar but more task-focused structure: one group aligns perfectly with the original demographics section~(M4); however, the original operational section has been decomposed into two focused analytical threads, one examining visit patterns~(M3) and another providing comprehensive patient volume breakdowns across both operational and demographic dimensions~(M1). Additionally, we identified a distinct analytical goal that more deeply explores the wait time and satisfaction scores based on whether or not patients were admitted~(M2), a relationship which had not been directly analyzed in the input dashboard.

{
\smaller
\begin{table}[]
\caption{For each input dashboard \textsc{use case}, we identify the key analytical goals and organize the charts into corresponding \textsc{groups} or mini-dashboards. This table also shows the number of \textsc{charts} and \textsc{data facts} ($original + new = total$) for each dashboard, as well as the number of new aggregated \textsc{insights} produced by our dashboard remixing pipeline.
}
\renewcommand{\arraystretch}{1.5}
\begin{tabular}{|l|c|c|c|c|}

\bottomrule
\textsc{use case}
& \textsc{groups} 
& \textsc{charts}
& \textsc{data facts} 
& \textsc{insights} \\

\hline
\begin{tabular}[c]{@{}c@{}} IMDB\end{tabular}
& 2
& 7 + 2 = 9
& \,\,53 + 7\,\,=\,\,\,60
& 8 \\

\hline
\begin{tabular}[c]{@{}c@{}} HR Attrition \end{tabular} 
& 1      
& 11 + 1 = 12                                                          
& \,\,98 + 3\,\,= 101   
& 4 \\   

\hline
\begin{tabular}[c]{@{}c@{}} Hospital ER \end{tabular} 
& 4      
& 11+17 = 28                                                          
& 103+89 =192   
& 12 \\

\hline
\begin{tabular}[c]{@{}c@{}} Call Center \end{tabular}  
& 3      
& 10 + 6 = 16                                                          
& 101+50 =151   
& 9 \\

\toprule
\end{tabular}

\label{table:examplestat}
\end{table}
}


\newcommand{\usecasetableORIGINAL}{
\begin{table}[]
\caption{\leilani{This table needs a caption, but otherwise I like it!}This table shows the number of \textsc{views} from the original dashboard, as well as the results from our dashboard remixing pipeline. For each dashboard, we identify the key analytical goals and organize the charts into corresponding \textsc{groups} or mini dashboards. \todo{finish caption} \jane{I wonder if \textsc{charts} should only show the NEW charts added? Should \textsc{facts} be split into original and NEW facts? Are there ever cases where an original view is NOT included in the final mini-dashboards?}}
\begin{tabular}{c|c|c|c|c|c}
\toprule
             & \begin{tabular}[c]{@{}c@{}}Views\end{tabular} & Groups & \begin{tabular}[c]{@{}c@{}}New \\ Charts\end{tabular} & \begin{tabular}[c]{@{}c@{}}Facts \\ (New/All)\end{tabular} & \begin{tabular}[c]{@{}c@{}}Aggregated \\ Insights\end{tabular} \\

\hline
\begin{tabular}[c]{@{}c@{}} IMDB \\ (D1)\end{tabular}        & 7                                                           & 2      & 2                                                           & 7/60    & 8                                                              \\
\hline
\begin{tabular}[c]{@{}c@{}}HR Attrition \\ (D2)\end{tabular} & 11                                                          & 1      & 1                                                          & 3/101   & 4  \\   
\hline
\begin{tabular}[c]{@{}c@{}}Hospital ER \\ (D3) \end{tabular} & 11                                                          & 4      & 17                                                          & 89/192   & 12                                                             \\
\hline
\begin{tabular}[c]{@{}c@{}}Call Center  \\ (D4) \end{tabular}  & 10                                                          & 3      & 6                                                          & 50/151   & 9                                                              \\
\hline
\end{tabular}
\label{table:examplestat}
\end{table}
}




\parHeading{The quantitative results} in \autoref{table:examplestat} show the pipeline behavior for different dashboard complexities and domains.  
The count of remixed \textsc{groups} reflects more about the analytical breadth of the original dashboard, rather than the original number of \textsc{charts}.
These results directly reflect the goal optimization process wherein our MILP algorithm seeks to find the minimum set of goals that maximize coverage of existing views. 

The number of groups is also affected by the \textit{density} of the input dashboard, which is related not only to the number of original charts, but also the number of data attributes and their relationships within and between views.
{See the supplemental material for a deeper reflection on how the number and use of data attributes across views impacts the grouping.}


New facts directly come from view enrichment, suggesting that dashboards with cross-attribute analytical gaps gain more from our enrichment approach. 
The number of aggregated \textsc{insights} is difficult to predict since it is based on the occurrence of patterns in the data and the data facts. 

\subsection{Expert Interview Design}
\label{sec:interview}

\begin{table}[h]
\centering
\caption{Expert interview participant backgrounds.}
\label{tab:participants}
\small
\begin{tabular}{llll}
\toprule
\textbf{ID} & \textbf{Role} & \textbf{Domain} & \textbf{YoE} \\
\midrule
E1 & Vis. researcher & BI software  & $>$ 20 \\
E2 & Graduate vis. researcher & Academia & 3 $-$ 5 \\
E3 & Vis. researcher \& engineer & Technology & 5 $-$ 10 \\
E4 & Vis. researcher \& engineer & BI software  & 5 $-$ 10  \\
E5 & Data scientist & Technology & 5 $-$ 10  \\
E6 & Data analyst & Finance  & 5 $-$ 10  \\
\bottomrule
\end{tabular}
\end{table}

{We conducted expert interviews with six dashboard experts (E1-E6, four women and two men; Table~\ref{tab:participants}). E1–E4 are visualization researchers and engineers with combined academic and industry experience who engage with dashboards from a design and evaluation perspective; E5–E6 are data practitioners who use dashboards professionally as analytical consumers. This mix allows us to assess both the pipeline's design quality and its practical utility. 
Each interview session was conducted remotely within 1.5 hours with screen sharing and think-aloud protocols. 
We instructed the experts to evaluate and compare the original dashboard and our pipeline result. We selected the more complex dashboards D3 and D4, each analyzed by three participants (E1/E4/E5 and E2/E3/E6, respectively). }

{When creators step back from their work, they try to read their dashboard as an outside viewer would; our pipeline aims to provide such an outside reading computationally. Our study examines whether this reading holds up against that of skilled human readers. In the first phase, each expert reviews the original dashboard and summarizes its main goals and a key takeaway, forming an independent reading that serves as the reference for the rest of the session. In the second phase, experts compare the pipeline's output against their own reading, noting where the groupings, enriched charts, and insights agree with it, where they diverge, and where they surface goals or patterns the expert had not identified. Finally, experts described how they would refine the dashboard in light of this comparison, acting as creators with the intent they externalized in the first phase. We did not provide experts with the creators' intended goals, both because these were unavailable for the public dashboards we used and because doing so would have removed the independent reading against which the pipeline's output is compared.}

\parHeading{Procedure.} After an introduction and their verbal consent, the experts were asked to \textbf{review the original dashboard} and complete two tasks: (1) summarize the main goals they perceived from the dashboard, focusing on what analytical questions it was designed to help users answer, and (2) identify one important takeaway. The experts were encouraged to think aloud during this process while we observed their navigation patterns and interpretation strategies. Next, we provided a brief introduction of our interface, and let the experts \textbf{explore the output of our dashboard remixing pipeline} for ten minutes. 

During this phase, we asked the experts to separately assess the grouping of charts, added visualizations, and the generated insights. 
Throughout the semi-structured interviews, the experts provided both quantitative ratings on 7-point Likert scales and qualitative feedback for the following questions: 
\begin{itemize}[leftmargin=*]
    \item How easy was it to identify the original dashboard's \textbf{goal}?
    \item  How easy was it to summarize the original \textbf{takeaways}?
    \item How satisfied are you with the new dashboard \textbf{grouping}? 
    \item How necessary are the new (\textbf{enriched}) charts for each goal?
    \item How satisfied are you with the content of the \textbf{insights}? 
\end{itemize}

\subsection{Expert Feedback}
\likert
We now report interview findings including expert responses to the Likert-scale
questions in Fig.~\ref{fig:likert}, qualitative feedback, and several opportunities for improvement or future work. 

\parHeading{Original Dashboard Goal and Takeaways.} Most participants found the original dashboards easy to interpret by \myquote{making educated guesses} based on labels and domain knowledge (E1, E3 and E4), but noted dashboard limitations in analytical depth. 
E6 hesitated in the absence of textual guidance or introductory context.
E1 stated that it was \myquote{pretty easy if you believed the objective is simply to report out the data}. 
E2 commented that each of the individual charts corresponded to clear analytical questions they can answer, but lacked additional cross-dimensional analysis. 
All experts identified basic data patterns from the original dashboards, and either chose the most statistically significant pattern such as an outlier (E3-E6) or the most relevant ones, assuming their roles and tasks based on the domain (E1 and E2).
E1 commented that \myquote{if you're trying to understand when the ER is busy, you can see the two really busier days. But if you then wanted to make any generalization about when things are busy and not. I'm having trouble looking for a bigger pattern.}
Similarly, E3 described the findings as \myquoteparticipant{mostly descriptive}{E3}, although they felt confident about what was visually presented. 


\parHeading{Remixed Grouping.} The experts found our grouping helpful for organizing the analysis and the main messages (ratings range from 4 to 7). 
E3 noted that \myquote{it breaks down all the results into relevant pieces} and provides guidance throughout the analytical process. 
When reviewing the call center overview dashboard, E2 appreciated how the grouping addressed different user needs: \myquote{if I'm arranging people's work time, I'll be more interested in the first [analytical goal]. But if I'm trying to improve the call center service, I can focus on the second part}. This representation can help dashboard designers better anticipate the potentially varied goals of their future viewers, and thus refine their designs to better realize those goals. For example, E2 and E4 were inspired to come up with more analyses that they would be interested in exploring; similarly, E3 began to \myquote{wonder what else is there?}  
 
\parHeading{Enriched Charts and Insights.}
Experts praised the inclusion of newly added charts for providing additional context.
For instance, E1 noted that one of the charts added to D3, the Hospital ER dashboard, \myquote{is the only one that really shows how much smaller the non-conforming genders are than anywhere else} (Fig.~\ref{fig:pipeline}\,\textcircled{\footnotesize A}, Chart 31). Similarly, E3 found a few of the added charts for D4, the call center dashboard, showed new combinations of attributes \myquote{that are more helpful compared to what was in the original dashboard}, as she pointed to the second mini-dashboard that compares sentiment and channel. 

Experts also acknowledged that the automated insights saved them time and effort, and uncovered overlooked patterns (E1, E2).
For example, E2 noted that \myquote{it summarized what I manually get and also more new ones that I didn't think of}.
While E4 appreciated the inclusion of new statistical values that could not be read directly from the charts, 
E1 claimed that \myquote{it was probably not interesting to a typical person using this kind of dashboard}.
By surfacing these values in our pipeline, designers can reflect on the utility for their own audiences and adapt the original dashboard designs accordingly.

\parHeading{Reflections and Refinements.}
When reviewing the original dashboard and results of our pipeline, experts identified several opportunities for refinement beyond the scope of our current implementation. For example, four experts (E1, E3-E5) suggested enhancing the charts, such as by using normalized values or percentages rather than raw counts in the stacked or grouped bar charts to make the comparison more salient.

When no outstanding pattern dominates the data, we observe that it becomes a matter of individual judgment and designer intent to determine which narrative to pursue. 
Our pipeline helps designers tell their own story by offering computational support based on heuristic-driven best practices, which is distinct from LLM-based systems that attempt to personalize narratives automatically. 
Designers retain full control over their own personalization and storytelling choices within an objectively grounded framework informed by established dashboard design principles. 
E3 also expressed a desire for more interactive control to explore other possible goals and chart combinations, while E4 wanted the ability to continue specific analyses within the same interface.

In terms of the insights themselves, E1 and E3 also noticed that, while accurate, some insights are less meaningful for well-known domains or other experts with sufficient prior knowledge. 
Two participants (E5 and E6) specifically noted that they found 
visually obvious insights unnecessary to state textually, preferring instead to read insights that are less discoverable by simply observing the charts. 
Although our semantic network is capable of specifying simple semantic relationships and domain knowledge, this observation suggests that the inclusion of more efficient ways to collect this knowledge and incorporate customized or learned scoring would be valuable for future instantiations of this approach. 
The limitations of automated fact mining and extraction are persistent challenges widely recognized in previous visualization and NL generation research~\cite{ding2019quickinsights}. Regardless of advances in statistical analysis and pattern detection algorithms, extracted facts often suffer from missing context, audience mismatch, and practical significance thresholds. Rather than attempting to solve the fundamental fact generation problem, our approach focuses on the orthogonal challenge of consolidating and summarizing patterns regardless of the fact quality.

\section{Discussion and Future Work}
\label{sec:discussion}

Our pipeline identifies predefined analytical goals, but more subtle, flexible, or overlapping objectives from complex dashboards remain challenging to infer.
{When the inferred goals diverge from a creator's intent because of these template limitations rather than gaps in the dashboard itself, the reflection becomes less informative, as E3's request for more control over which goals are explored suggests.}
Additionally, the current approach applies universally to all users, thereby maintaining consistency but overlooking individual differences in preferences, expertise, and users' specific analytical needs. 
Furthermore, exploring a variety of other domains like finance, healthcare, and retail would help identify new domain-specific challenges to refine the goal inference algorithm.

Generating a complete, analytically coherent dashboard directly from a stated goal remains an open research problem. A stated goal does not uniquely determine a dashboard: many different sets of views and groupings could satisfy the same high-level intent. Existing goal- or NL-conditioned recommendation systems, including Medley~\cite{pandey_medley_2022} and Bolt~\cite{srinivasan_bolt_2023}, address the initial mapping from intent to candidate views, but the problem of iteratively refining toward the one dashboard a creator actually wants, among the many that would technically satisfy the goal, remains open. Our pipeline takes a different angle on dashboard authoring: rather than generating views from a goal, it takes an already-authored dashboard, however it was produced, and helps creators understand and reorganize what it currently expresses. This angle remains useful independent of progress on generation, since a dashboard's expressed content drifts from its original goal as creators iterate and refine their focus, regardless of whether its views were manually authored or generated. The pipeline is thus complementary to generation-based approaches rather than an alternative solution to the same problem. 

To enrich charts and insights based on the inferred goal, we currently focus more on the data relationships than the visual representation since chart recommendation plays a limited role in our pipeline. 
More sophisticated visualization recommendation approaches leveraging perceptual and task-based effectiveness principles could improve presentation clarity, enhance understanding, and make hidden data patterns more salient through appropriate data and visual transformations.

Our rule-based default reflects a deliberate design choice for the reflection use case to offer traceable outputs so that users can verify or contest the pipeline's feedback~\cite{amershi2019guidelines, heer2019agency}. Rule-based computation satisfies this as fact extraction and pattern detection are established in prior work~\cite{ding2019quickinsights,Srinivasan2019Voder} and formally specified. 
While our data fact extractors recognize and generate insights with high fidelity, they would benefit from additional domain knowledge and contextual understanding. LLM-generated text optimizes for fluency and apparent coherence but does not guarantee that stated patterns are grounded in the underlying data. 
That said, a rigorous comparison of rule-based and LLM-based approaches would require human annotation at scale and a carefully engineered LLM baseline, which is beyond the scope of this paper. To explore where LLM integration adds value within our architecture, we implemented optional LLM-based enhancements at each pipeline stage (Section~\ref{sec:system}), including an agentic tool-selection mode, a display-only description rewriter, and two narrative aggregation modes. LLMs and VLMs could also play a deeper role in future work by enriching goal inference with domain-specific context, enabling more flexible relationship detection, or personalizing scoring parameters from user feedback. 

{Our dashboard specification captures a static snapshot of a dashboard's state, not its full interactive behavior. As a result, three aspects of dashboard interactivity fall outside the current scope. First, viewer-applied filters at runtime change what a view displays, but any patterns visible only within a filtered subset go uncaptured by the current implementation. Second, view coordination such as linked selections or brushing is a signal of relatedness that creators often build in deliberately, but the pipeline does not extract or use it. Third, spatial layout (\eg{} the proximity, grouping, and relative sizing of views) encodes further editorial judgment that is not currently used in goal enumeration or selection. Incorporating interaction traces, coordination graphs, and layout as additional inputs to goal enumeration are promising directions for future work. }

{
A limitation of our evaluation is that experts' intents were formed by reading dashboards authored by others, rather than arising from their own analysis of the data. A creator's intent typically develops over the course of exploration and may be more specific or more strongly held. In addition, experts proposed refinements but did not carry them out and re-run the pipeline. A study in which creators apply the pipeline to their own dashboards across multiple rounds of authoring would show how its reflections inform refinement over time.
}
\section{Conclusion}
In this paper, we contribute a pipeline to facilitate dashboard design reflection.
Our pipeline aims to automatically enrich and restructure creators' dashboard designs by first inferring their analysis \emph{goals}, before generating new views and insights to complement the original analysis.
Our approach then leverages an insight aggregation method to concisely summarize the data \emph{insights} associated with each goal in natural language. 
We demonstrated four use cases using real-world dashboards, and observed the effectiveness of the pipeline as a dashboard design aid from interviews with six design experts. 

\section*{Acknowledgments}
This research was supported in part by the NSF (award \# 2141506, \# 2402718), MotherDuck, and Google. This work originated as a research internship at Adobe.      

\bibliographystyle{IEEEtran}
\bibliography{sections/bibliography}

\vspace{-1cm}

\begin{IEEEbiographynophoto}
{Junran Yang} is a PhD student at the University of Washington. Her research interests are visualization toolkits and interactive systems, with a focus on dashboard authoring and understanding.
\end{IEEEbiographynophoto}
\vspace{-1cm}

\begin{IEEEbiographynophoto}
{Leilani Battle} is the Robert E. Dinning Endowed Associate Professor in Computer Science \& Engineering at the University of Washington. Her research focuses on developing interactive systems for exploring large, complex datasets. She earned a PhD degree in Computer Science from MIT (2017). 
\end{IEEEbiographynophoto}

\vspace{-1cm}
\begin{IEEEbiographynophoto}%
{Jane Hoffswell} is a senior research scientist at Adobe, specializing in information visualization and human-computer interaction. 
Her research generally explores the design of responsive data visualizations, interactive systems, as well as data-driven visual storytelling.
Jane received her Ph.D. in Computer Science from the University of Washington (2020). 
\end{IEEEbiographynophoto}

\end{document}